\documentstyle[preprint,aps]{revtex}
\tighten
\begin{document}
\draft
\title{Dynamical Ambiguities in Singular Gravitational Field}
\author{George Jorjadze}
\address{Department of Theoretical Physics, Razmadze Mathematical
Institute, Tbilisi, Georgia.\\}
\author{W{\l}odzimierz Piechocki}
\address{Field Theory Group, So{\l}tan Institute for Nuclear 
Studies, Warsaw, Poland.\\}
\date{\today}
\maketitle

\begin{abstract}
We consider particle dynamics in singular gravitational field.
In 2d spacetime the system splits into two independent
gravitational systems without singularity. Dynamical integrals of
each system define $sl(2,R)$ algebra, but the corresponding symmetry
transformations are not defined globally. Quantization
leads to ambiguity. By including singularity one can 
get the global $SO(2.1)$ symmetry. 
Quantization in this case leads to unique quantum theory. 
\end{abstract}

\pacs{11.30.Na, 02.40.Ky, 02.20.Sv}

In the Einstein theory of gravity singularities 
play a fundamental role. Since the four dimensional quantum gravity
is still under construction, it makes sence to investigate
spacetime singularities at the level of 2-dimensional theory
to get some insight.

Let us consider a relativistic 
particle of mass $m_0$ moving in the gravitational field 
$g_{\mu \nu}(X)$ ($X:=(x^0,x^1)$; $(\mu ,\nu = 0,1))$.
The action describing such a system is proportional to
the length of a particle world-line and reads
\begin{equation}
S= - m_0\int d\tau 
\sqrt{g_{\mu \nu}(X(\tau))\dot{x}^{\mu}(\tau)\dot{x}
^{\nu}(\tau}),
\end{equation}
where $\tau$ is an evolution parameter along the 
trajectory $x^{\mu}(\tau)$, and 
$\dot{x}^{\mu}:= dx^{\mu}/d\tau$.
		     
One can always choose such local coordinates on a 2-dimensional 
pseudo-Riemannian manifold that 
$g_{\mu \nu}(X)$
takes the form [1] 
\begin{equation}
 g_{\mu \nu}(X) = \exp{\phi (X)}\;\;\left( \begin{array}{cr}
 1 & 0 \\ 0 & -1
 \end{array} \right),
 \end{equation}
where $\phi (X)$ is a field.

It is known that the Einstein-Hilbert action in 2-dimensions
does not lead to dynamical equations for the metric tensor 
$g_{\mu \nu}$ (i.e., for the field $\phi (X))$.
The Liouville field equation
\begin{equation}
(\partial ^2_0 - \partial ^2_1)\;\phi (x^0,x^1) + R_0 \exp{\phi 
(x^0,x^1)} = 0,
\end{equation}
(where $R_0$ is a non-zero constant) is usually considered
as a model of $2d$ gravity \cite{1,2} and it describes a spacetime 
manifold with a
constant curvature $R_0$.
One can show that the equation $R_{\mu\nu}-(1/2)R_0~g_{\mu\nu}=0$
(where $R_{\mu\nu}$ is a Ricci tensor) in the conformal gauge (2) is
equivalent to (3).
 
In recent paper [3] we  have investigated the mathematical 
aspects of particle dynamics in an arbitrary Liouville field. 
We have shown that the particle dynamics for different Liouville 
fields looks locally the same due to the local $SO(2.1)$ symmetry.
The aim of this Letter is to investigate the dynamical ambiguities
for a particle motion 
in a singular Liouville field
\begin{equation} 
{\phi}(t,x) =
-2\ln (m|t|),~~~~~\mbox{with}~~~~~~m:=\sqrt{-R_0/2}~~~~(R_0<0),
\end{equation}
given on the plane ($t$, $x$), where $t:=x^0$ and $x:=x^1$.

We interpret $t$ and $x$ as time and space coordinates,
respectively. Particle trajectories $x=x(t)$ are time-like
$(|\dot x|<1)$. In our units, the `velocity of light' is 1 and 
$\hbar =1$.

The Liouville field (4) leads to the singularity of the spacetime 
metric (2). The proper time of the particle from $t_0$
to $t$ ($t_0 < t < 0$)
$$
\int_{t_0}^{t}dt'~\sqrt{g_{00}(t',x(t'))} =
 \int_{t_0}^{t}\frac{dt'}{m|t'|}
$$
divergies for $t\rightarrow 0$, i.e., particle needs infinite
(proper) time to reach the singularity. 
The same situation is for the interval
$(t , t_0)$, with ($0<t<t_0$) and $t\rightarrow 0$. 
Therefore, the dynamics of the particle can be
considered for $t<0$ and $t>0$ separately. In such interpretation
we deal with two independent dynamical systems without singularities.
 The proper time for each system goes from $-\infty$ to $+\infty$.
 The coordinate $t$ can be considered as a time coordinate of some
 `outside observer' in three (or higher) dimensional spacetime.
Since the dynamics does not depend on the 
choice of coordinates we can use $t$ as the time coordinate
for each system. We denote these systems by ${\cal {S}}_+$ 
(for $t>0$) and 
${\cal {S}}_-$ (for $t<0$).

The Lagrangian of (1) with the Liouville field (4) 
in the gauge 
$t-\tau =0$ reads
\begin{equation}
L=-\frac{a}{|t|}\sqrt{1-\dot x^2},~~~~~~~
(a:=\frac{m_0}{m}).
\end{equation}

The dynamics for both systems ${\cal {S}}_\pm$ is similar
and we use the same
notations. In the case of differences we specify the
corresponding formulae using the parameter $\epsilon :=t/|t|$.

Since $L$ is homogeneous in space, particle momentum $P=\partial L/
\partial \dot{x}$ is conserved. There are two other dynamical 
integrals 
\begin{equation}
K=Px - E t,~~~~~~~M=P(t^2+x^2)-2txE,
\end{equation}
where $E=\sqrt{P^2+(a/t)^2}$ is the energy of the particle.
$E$ is not conserved since $L$ depends on time. The conservation of 
$K$ is connected with the dilatation symmetry $(t\rightarrow \lambda t
, x \rightarrow \lambda x,~ \lambda >0)$ of our systems. 

The infinitesimal symmetry transformations 
 of the particle trajectories for the dynamical integrals
$P$, $K$ and $M$ are
$$
x(t) \rightarrow x(t)+\varepsilon_1,~~~~~~ 
x(t) \rightarrow x(t)+\varepsilon_2[x(t)-t\dot x(t)],
$$
$$
x(t) \rightarrow x(t)+\varepsilon_3[t^2+x^2(t)-2tx(t)\dot x(t)],
$$
respectively. It is clear that the first two transformations can 
be defined globally
for each system ${\cal {S}}_\pm$, 
while the third transformation needs further investigation.

The dynamical integrals (6) define trajectories of the particle. 
For $P=0$ we get
\begin{equation}
K=-\epsilon a,~~~~~~~~~~x=-M\epsilon/2a~~~~~~~~~~~(\epsilon
= t/|t|),
\end{equation}
which describes a rest particle. For $P\neq 0$ the trajectories are
hyperbolas 
\begin{equation}
x=\frac{K+\epsilon \sqrt{P^2t^2+a^2}}{P},
\end{equation}
with the light-cone asymptotics (when $t\rightarrow\pm\infty )$
and with  zero velocity when $t\rightarrow 0$.

In Hamiltonian formulation $P$ and $x$ are canonically conjugated 
coordinates, $\{P,x\}=1.$ The commutation relations of the dynamical 
integrals $P,K $ and $M$ define the  algebra $sl(2,R)$
\begin{equation}
\{P,K\}=P,~~~~~~~~~\{K,M\}=M,~~~~~~~~~~\{P,M\}=2K.
\end{equation}
Using (6) we get the following relation 
\begin{equation}
K^2-PM=a^2.
\end{equation}
The hyperboloid (10) in $(P,K,M)$ space
is the coadjoint orbit of $SL(2,R)$ group [4]. 
 The Poisson brackets (9) define the symplectic form on the hyperboloid 
(10), which has the global $SO(2.1)$ symmetry generated 
by (9).
According to (7) and (8) each point ($P, K, M$) of the hyperboloid 
(10) specifies
the trajectory uniquely (for each system separately). 
Thus, we can associate
the set of trajectories with the corresponding set of points on the
hyperboloid. 
Let us check whether all points of the hyperboloid (10) specify 
the dynamics.
For a given $t$, Eq.(6) defines the map from the $(P,x)$ plane to the
hyperboloid (10). For $t<0$ this map covers the whole hyperboloid 
except the line given by ($P=0$, $K=-a$). Similarly, for $t>0$ 
the line ($P=0$, $K=a$) is not available. Therefore, none of the 
systems separately have the global $SO(2.1)$ symmetry. 

Here it is convenient to introduce a new time independent coordinate 
$Q_\epsilon$ 
(canonically conjugated to $P$)
\begin{equation}
Q_\epsilon=(K+\epsilon a)/P.
\end{equation}
The plane ($P, Q_-$) is isomorphic to the hyperboloid (10) without 
the line ($P=0$, $K=-a$)
and it describes the space of all trajectories for $t<0$.
Similarly, the space of trajectories for the system ${\cal {S}}_+$ is 
given by the plane ($P, Q_+$).
The dynamical integrals $P$ 
and $K$ generate global symmetry transformations on each 
($P, Q_\epsilon$) plane, 
while the transformations generated by $M$ are defined only 
locally. 

Now we quantize our system. 
The coordinates ($P,Q_\epsilon$) are convenient for the canonical
quantization since the dynamical integrals are  linear in $P$ 
\begin{equation}
K=PQ_\epsilon -\epsilon a,~~~~~~~
M=PQ_\epsilon^2 + 2\epsilon a Q_\epsilon
\end{equation}
and the operator ordering problem can be easily solved [3]. 
For the corresponding operators we get
\begin{equation}
\hat{P}=-i\partial_{Q_\epsilon},~~~~~
\hat{K}=-iQ_\epsilon\partial_{Q_\epsilon} -\epsilon a - i/2,~~~~~~
\hat{M}=-iQ_\epsilon^2\partial_{Q_\epsilon}-(i+2\epsilon a)Q_\epsilon.
\end{equation}
 
These operators are Hermitian on $L_2(R)$ and give the representation 
of $sl(2,R)$ algebra. However, according to the quantization 
principle quantum observables should be represented by self-adjoint
operators [5]. The operators $\hat{P}$ and $\hat{K}$ have 
unique self-adjoint extensions, while the self-adjoint extension of 
$\hat{M}$ is non-unique [3]. This ambiguity is parametrized by a 
complex number $z$ of unit norm ($\mid z\mid=1$). Therefore, we have a 
continuous set of unitary non-equivalent quantum systems which 
describe unitary non-equivalent representations of the universal 
covering group $\overline{SL(2,R)}$ [6]. 
Note that for $z=1$ we have the unitary irreducible representation
of $SO(2.1)$ group.
The quantum ambiguity 
here is connected with a lack of the global $SO(2.1)$ symmetry
of both systems ${\cal {S}}_\pm$.

Now, we consider another aproach. For the outside observer both
systems ${\cal {S}}_\pm$ are two parts of a one system with
$t$ (time) going continuously from negative to positive values.
Such a system has singularity at $t=0$ and we should specify the
way of `glueing' trajectories for $t<0$ and $t>0$.

Using (8) and (11) we get
$$
\lim_{t\rightarrow \pm 0}x(t)=Q_\pm,~~~~~~\lim_{t\rightarrow \pm 0}
\dot x(t)=0.
$$
Thus, the velocity is continous at $t\rightarrow 0$ and continous 
trajectory
implies $Q_+=Q_-$. On the other hand, if the dynamical integrals
$P, K$ and $M$ are conserved (when the particle
`passes' the singularity) we have (see (11))
\begin{equation}
Q_+-Q_-=\frac{2a}{P}.
\end{equation}
Therefore, conservation of $P, K$ and $M$ numbers 
and continuity of trajectories
are incompatible. In the case of continous trajectories 
only one integral can be
conserved. For example, if $P$ is conserved and trajectory 
is continuous we have
\begin{equation}
x(t)=\frac{K_--\sqrt{P^2t^2+a^2}}{P},~~\mbox{for}~~t<0;~~~~~~
x(t)=\frac{K_++\sqrt{P^2t^2+a^2}}{P},~~\mbox{for}~~t>0; 
\end{equation}
with $K_--K_+=2a$. Note that such trajectory is not smooth since 
$\ddot x(t)$
has discontinuity at $t=0$. The smooth trajectories correspond to 
change of 
sign of all integrals $P, K$ and $M$ when the particle passes 
the singularity
(see (8)).  There are many different ways of glueing trajectories. 
Each way corresponds to some identification of the points on 
$(P_+, Q_+)$ and $(P_-, Q_-)$ planes and defines the set of trajectories
$x(t)$ ($-\infty <t<+\infty$).
Choosing some definite rule of identification we obtain the set of all
trajectories  for the outside observer. 
For example, the identification ($P_+=-P_-,~ Q_+=Q_-$) 
leads to the smooth
trajectories, while the trajectories (15) correspond to 
($P_+=P_-,~ Q_+=Q_-$).

Let us choose the set of trajectories in such a way that all dynamical 
integrals $P$, $K$ and $M$ are conserved 
(when the particle passes the singularity)
and the set of all trajectories
for the whole system has the global $SO(2.1)$ symmetry. As it 
was mentioned
above, the conservation of all dynamical integrals leads to the 
discontinuity
of trajectories at $t=0$.
We interpret these discontinuities in the following way:
At $t=0$ the particle with the momentum $P\neq 0$ is `annihilated' 
at $x=(K-a)/P$ and `created' at 
$x=(K+a)/P $ by the `spacetime singularity'. 
For $P=0$ and $t<0$ we have $K=a$. Such a particle is 
annihilated and it cannot appear for $t>0$, since 
for $t>0$ there are no trajectories with $P=0$ and $K=a$. For $P=0$ 
and $t>0$ there are trajectories with only $K=-a$ and there are no 
`corresponding' trajectories for $t<0$. Such a particle is created 
by the singularity. 

Thus, we get the following set of `trajectories':
\begin{description}
\item[(i)] $P\neq 0$, $K$ is arbitrary, trajectories (8) with 
discontinuity
$2a/P$  at $t=0$ (see (14)); 
\item[(ii)] $P=0$, $K=a$, $M$ is arbitrary, $x=M/2a$,
for $t<0$ and there is no trajectory for $t>0$; 
 \item[(iii)] $P=0$, $K=-a$, $M$ is arbitrary, trajectories start at 
$t=0$ and $x=-M/2a$ for  $t>0$.
\end{description}

The set of trajectories defined by (i) and (ii) corresponds to the
$(P, Q_-)$ plane, which is isomorphic to the hyperboloid (10)
without the line $(P=0,~ K=-a)$. Completeing this set of trajectories
by (iii) we cover the entire hyperboloid (10). In this way we arrive at
the global $SO(2.1)$ symmetry in the space of all trajectories 
with the conservation of all dynamical integrals.

To quantize the system we use the
following parametrization of the hyperboloid (10)
\begin{equation}
P=J(1-\cos\varphi )-a\sin\varphi ,~~~K=-J\sin\varphi +a\cos\varphi ,~~~
M=J(1+\cos\varphi )+a\sin\varphi ,
\end{equation}
where $(J,\varphi )$ are the cylindrical coordinates ($J\in R$,
$\varphi\in S^1$). One can show that (16) gives the unique 
parametrization
of the hyperboloid  and the Poisson brackets
(9) are equivalent to the canonical commutation relations 
$\{ J,\varphi \}
=1$. The dynamical integrals (16) are linear in $J$ and it again
simplifies the operator ordering problem. Applying the canonical 
quantization
rule $\hat J=-i\partial_\varphi$, we get the
following operators
$$
\hat P=-i(1-\cos\varphi )\partial_\varphi-(a+i/2)\sin\varphi ,~~~~~~
\hat K=i\sin\varphi \partial_\varphi+(a+i/2)\cos\varphi ,
$$
$$
\hat M=-i(1+\cos\varphi )\partial_\varphi+(a+i/2)\sin\varphi.
$$
These operators are self-adjoint on $L_2(S^1)$ and define the unitary
irreducible representation of $SO(2.1)$ group. 
This representation is unitarily equivalent to the above mensioned 
representation of $\overline{SL(2,R)}$ group for $z=1$.

Thus, taking the spacetime manifold to be 
$R^2=\{ (t, x)~|~t\in R,~x\in R\}$
we can get 
the global $SO(2.1)$ symmetry 
of the system. However, the spacetime has now the singularity.
 This results 
in a strange phenomena  at the classical level
(particle creation and annihilation by 
spacetime singularity), but gives the unique 
quantum theory.

Note that  one can also join both
systems ${\cal {S}}_+$ and ${\cal {S}}_-$ into a one system with 
the global $SO(2.1)$ symmetry and without spacetime singularity, 
but with different spacetime topology. 
It can be achieved by a map from the 
half planes $(t,x;~t<0)$  and $(t,x;~t>0)$ to the hyperboloid
$y_1^2+y_2^2-y_0^2=r^2$ ($r:=1/m$). This  hyperboloid has a constant
curvature in three dimensional Minkowski space and is invariant
under the corresponding $SO(2.1)$ transformations.
The map is defined by 
\[ y_0=\frac{t^2-x^2-r^2}{2t},~~~~~y_1=\frac{t^2-x^2+r^2}{2t},~~~~~~
 y_2=\frac{rx }{t},
 \]
where $y_0$ is time coordinate.
This map covers the entire hyperboloid except two lines:
($ y_1=y_0,~y_2=r$) and ($ y_1=y_0,~y_2=-r$). Thus, the range
of this map should be completed by these two lines to get the
entire hyperboloid.

It is clear that such description corresponds to the choice
of spacetime manifold
to be a hyperboloid.

This work was supported by the grants from:
INTAS, RFBR (96-01-00344),
the Georgian Academy of Sciences, the
Polish Academy of Sciences, and the 
So{\l}tan Institute for Nuclear Studies.

\end{document}